\documentclass[12pt]{article}
\usepackage{graphicx}
\usepackage{amssymb}
\usepackage{xspace}
\usepackage{multirow,booktabs}

\def\etal{{\it et al.}\xspace}

\begin{document}
\begin{titlepage}
\begin{center}
\Huge {\bf Comparison of available measurements of the absolute air-fluorescence yield and determination of its global
average value}
\par
\vspace*{2.0cm} \normalsize {\bf {J. Rosado, F. Blanco, F. Arqueros}}
\par
\vspace*{0.5cm} \small \emph{Departamento de F\'{i}sica At\'{o}mica, Molecular y Nuclear, Facultad de Ciencias
F\'{i}sicas, Universidad Complutense de Madrid, E-28040 Madrid, Spain}
\end{center}
\vspace*{2.0cm}

\begin{abstract}
	Experimental results of the absolute air-fluorescence yield are given very often in different units (photons/MeV or
photons/m) and for different wavelength intervals. In this work we present a comparison of available results normalized
to its value in photons/MeV for the 337~nm band at 1013~hPa and 293~K. The conversion of photons/m to photons/MeV
requires an accurate determination of the energy deposited by the electrons in the field of view of the experimental
set-up. We have calculated the energy deposition for each experiment by means of a detailed Monte Carlo simulation and
the results have been compared with those assumed or calculated by the authors. As a result, corrections to the
reported fluorescence yields are proposed. These corrections improve the compatibility between measurements in such a
way that a reliable average value with uncertainty at the level of 5\% is obtained.

\end{abstract}
\end{titlepage}

\section{Introduction}

The air-fluorescence yield $Y$, defined as the number of photons per unit of energy deposited by the shower in the
atmosphere, is a key calibration parameter which determines the energy scale of fluorescence telescopes. A number of
absolute measurements of the fluorescence yield have been carried out in laboratory experiments in recent years. In
these experiments a beam of charged particles crosses a collision chamber filled with air at known conditions,
generating fluorescence radiation which is measured by an appropriate optical system. Many
authors~\cite{Kakimoto,Nagano_04,Lefeuvre,MACFLY,AirLight} have used electrons from a source of $^{90}$Sr with energy
around 1~MeV. Other absolute measurements have been performed with higher energy electrons from
accelerators~\cite{MACFLY,FLASH_08}. Finally the AIRFLY collaboration has carried out accurate  measurements using a
120~GeV proton beam and a preliminary value is presented in these Proceedings~\cite{AIRFLY_nagoya}. In
table~\ref{tab:comparison} a summary of available measurements is shown.

The atmospheric fluorescence emission in the spectral range of interest, i.e., $\sim 300 - 400$~nm, basically comes
from the Second Positive (2P) system of N$_2$ and the First Negative (1N) system of N$^+_2$. The excitation cross
section of the 2P system, dominant at atmospheric pressure, peaks at about 15~eV decreasing strongly with an $E^{-2}$
dependence. As a consequence, high-energy electrons themselves are very inefficient for the generation of air
fluorescence. In fact, fluorescence emission along the track of an energetic charged particle is mainly induced by
low-energy secondary electrons ejected in successive ionization
processes~\cite{Blanco,Arqueros_Astropart_Phys,Arqueros_NJP}.

Comparison of the absolute values of the fluorescence yield is not trivial. For instance some authors measure single
intense fluorescence bands while others detect the integrated fluorescence in a wide spectral range. In this work we
have normalized the available measurements of the absolute fluorescence yield to their values for the 337~nm band using
experimental relative intensities~\cite{AIRFLY_P} in good agreement with theoretical predictions~\cite{Arqueros_NJP}.
Some measurements are given in units of ph/m, and converted by the authors into ph/MeV assuming that beam electrons
deposit all their lost energy in the field of view of the collision chamber. We have calculated the energy deposition
in these experiments using a simulation algorithm~\cite{Arqueros_NJP} which has provided us with the appropriate
corrections factors. Other experiments have determined accurately the energy deposition as well as the geometrical
factors by means of well known MC codes, e.g., GEANT4 and EGS4. In these cases our simulation predictions have been
compared with those of the corresponding experiments. In general, a reasonable agreement has been found, and thus, the
corresponding corrections factors that we propose, to be consistent with our simulations, are small.

Unlike other MC codes like GEANT or EGS, our simulation algorithm~\cite{Arqueros_NJP} has been developed to treat
individual interactions of both primary and secondary electrons with the molecules of the medium. All processes giving
rise to energy deposition are included. Molecular excitation for the emission of 2P and 1N photons are treated
separately in such a way that fluorescence emission  can be also calculated and therefore a theoretical fluorescence
yield can be obtained too. GEANT4 simulations of energy deposition in air, carried out by MACFLY~\cite{MACFLY} and
AIRFLY~\cite{AIRFLY_geant4} are in agreement with our calculations at the level of 2\% (1\%) for electron energies in
the GeV (MeV) range.

By means of this algorithm we have performed two kind of simulations, i.e., generic simulations where primary electrons
are forced to interact in the center of a sphere of radius $R$ filled with air at given pressure and detailed
simulations~\cite{Rosado_Astropart_Phys} including the geometry as well as other experimental features. By comparison
with the detailed ones we have found that even the generic simulations provide accurate results on energy deposition as
far as the radius of the sphere is a reasonable representation of the size of the experimental set-up. This behavior is
expected since energy deposition is a weak function of $R$~\cite{Arqueros_NJP}.

As a result of our calculations, normalized values of the air-fluorescence yield (ph/MeV) for the 337~nm band at
1013~hPa and 293~K have been obtained. We will compare below two set of values, i.e., those using the
assumptions/calculations of the authors (uncorrected values) and those resulting from our corrections (corrected
values). The results presented here are updated values of those previously shown in~\cite{Rosado_Astropart_Phys} after
some improvement in the simulation code.

\section{Simulation of fluorescence yield experiments}
We have performed a detailed simulation for the experiments of Nagano \etal~\cite{Nagano_04}, AirLight~\cite{AirLight},
FLASH-2008~\cite{FLASH_08} and MACFLY~\cite{MACFLY}. Experiments of Kakimoto \etal~\cite{Kakimoto} and Lefeuvre
\etal~\cite{Lefeuvre} have been also analyzed although a dedicated simulation was not carried out. In
table~\ref{tab:comparison} the main parameters of these experiments are displayed.

\begin{figure}[!b]
\centering
\includegraphics[width=0.65\linewidth]{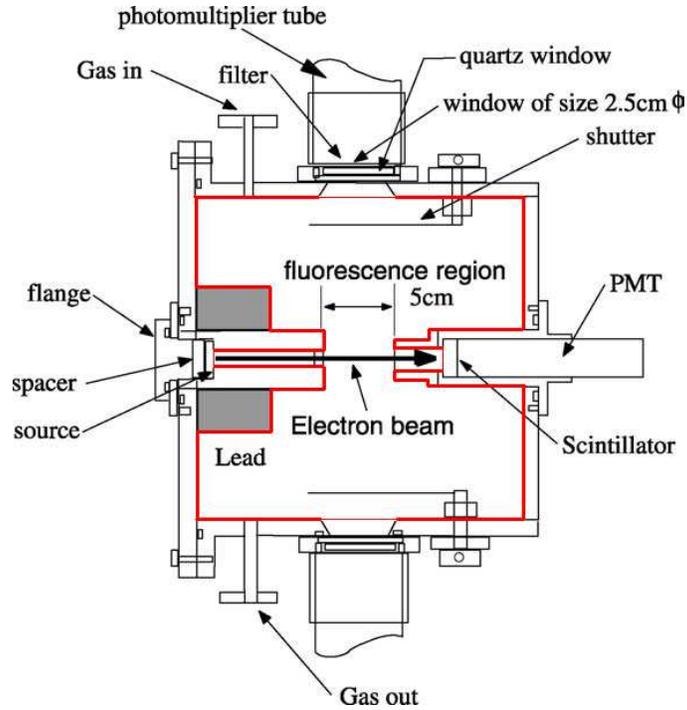}
\caption{Schematic drawing of Nagano \etal experiment (taken from~\cite{Nagano_03}). Electrons from a $^{90}$Sr source
are beamed and detected by a scintillator counter. The thick contour represents the geometry implemented in our
simulation.}
\label{fig:Nagano_chamber}
\end{figure}

The Nagano \etal experiment~\cite{Nagano_04} has been simulated including the geometry of the collision chamber (see
figure~\ref{fig:Nagano_chamber}) as well as other experimental features, e.g., electron-photon coincidence triggered by
an electron hitting the scintillator, the light-blocking effect of the collimation channel and the scintillator
enclosure. As input to the simulation, we have used a truncated gaussian fit to the measured $^{90}$Sr spectrum
of~\cite{Nagano_03}, with mean energy of 0.85~MeV, end point at about 2~MeV and a threshold of 0.3~MeV. According to
our simulation, most beam electrons of low energy are scattered away at atmospheric pressure and therefore the spectrum
of contributing electrons is shifted to higher energies, with a mean value of 1.11~MeV at atmospheric pressure. The
statistical uncertainty of this simulation was below 1\%.

The approximations made by Nagano \etal are the following. Firstly, they assumed that all the fluorescence is emitted
from the beam itself while a fraction of the light is produced by high-energy secondaries well outside the beam region.
Secondly, they calculated the number of photons per meter assuming they are emitted in a length equal to the gap
distance $\Delta x_{\rm gap}=5$~cm, and thus neglecting the dispersion of beam electrons. Finally, for the calculation
of the fluorescence yield, they assumed that the energy deposited in the observation volume equals the collisional
energy loss $({\rm d}E/{\rm d}x)_{\rm loss}$ at 0.85~MeV as given by the Bethe-Bloch formula. Therefore, the
fluorescence yield value $Y_{\rm Nag}$ reported by Nagano \etal, should be corrected by three factors accounting for
the above mentioned approximations:

\begin{equation}\label{eq:Nagano}
Y=Y_{\rm Nag}\frac{\Omega_{\rm beam}}{\Omega}\frac{\Delta x_{\rm gap}}{\Delta x}\frac{\left({\rm d}E/{\rm d}x\right)_{\rm
loss}}{\langle{\rm d}E/{\rm d}x\rangle_{\rm dep}}\,.
\end{equation}

From our simulation we have found that the acceptance correction, i.e., the $\Omega/\Omega_{\rm beam}$ increases the
fluorescence yield in about 1\%. However this effect is nearly compensated by that of the gap length (same size but
opposite direction). From our simulation, we have obtained the average energy deposited per electron and unit path
length $\langle{\rm d}E/{\rm d}x\rangle_{\rm dep}$ as the ratio between the integrated energy deposition and the total
path length of beam electrons within the observation volume giving a value somewhat smaller than $\left({\rm d}E/{\rm
d}x\right)_{\rm loss}$. As a result, according to our simulation, the fluorescence yield of Nagano \etal should be
increased by 6\%.

The result of the deposited energy from our detailed simulation of the Nagano \etal experiment is fully compatible with
the predictions of the generic simulation for a simple geometry, assuming a sphere of radius $R=5$~cm and an electron
energy of 1~MeV.

The experimental technique used in AirLight~\cite{AirLight} is similar to the one of Nagano \etal, but with a collisional
chamber larger by about a factor of two. In~\cite{AirLight} a detailed simulation of the experiment is carried out
using GEANT4, including electron scattering by elements of the chamber (e.g., collimator walls, scintillator). Both
beam electrons and secondary ones are thoroughly tracked allowing the authors to determine the geometrical acceptance
as well as the energy deposited in the observation volume.

For this experiment we have carried out a simulation including propagation of beam electrons and the geometrical
details of the set-up. The aim of our simulation was to calculate the integrated energy deposition as a function of the
electron energy to be compared with the results found by AirLight. While at low pressure we have obtained results in
fair agreement with those reported in~\cite{AirLight} (within 5\% at 50~hPa), at atmospheric pressure our simulation
predicts a larger energy deposition. For instance at 800~hPa, deviations range from about 5\% at low electron energy up
to 20\% at 2000~keV. The origin of this discrepancy is still unclear. These simulation results have been
analyzed~\cite{Rosado_Astropart_Phys} taking into account that AirLight reports fluorescence yield values extrapolated
at null pressure together with a set of quenching parameters. As a result we propose a correction of -7\% to the
AirLight fluorescence yield.

\begin{figure}[!b]
\centering
\includegraphics[angle=-90,width=0.65\columnwidth]{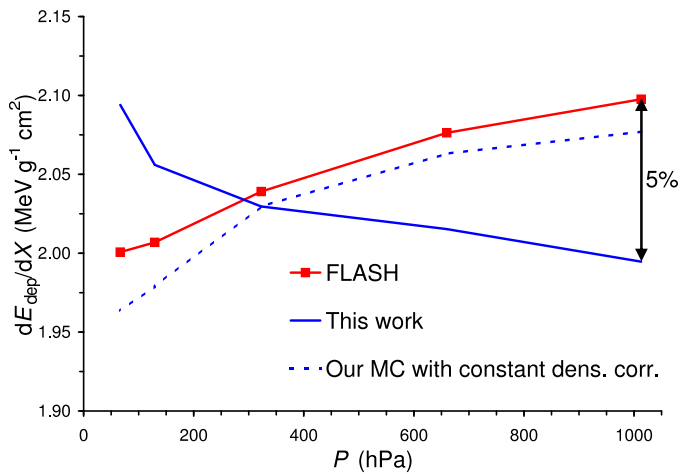}
\caption{Energy deposited per unit mass thickness as a function of
pressure for the FLASH experiment~\cite{FLASH_08}. The largest disagreement between our calculations and the simulations
of FLASH is 5\%. Note the different behavior of the pressure dependence. For comparison, results from our simulation
applying a constant density correction at 323 h Pa in the whole pressure range are also shown.}
\label{fig:FLASH_08}
\end{figure}

The FLASH experiment~\cite{FLASH_08} uses an intense beam of 28.5~GeV electrons from a linear accelerator. We have
carried out a detailed simulation to evaluate the energy deposited in the chamber as well as the average geometrical
acceptance. Statistical uncertainties were reduced to less than 0.3\%. In our algorithm, the observation volume has
been approximated by a rectangular prism of dimensions 1.60~cm$\times 1.67$~cm$\times 72$~cm. In
figure~\ref{fig:FLASH_08}, the resulting values of energy deposition are shown as a function of pressure together with
those obtained by FLASH using EGS4 (details in~\cite{Rosado_Astropart_Phys}). At low pressures ($\sim 300$~hPa), both
results are in very good agreement while at high pressure the discrepancy is of about 5\%. Notice the different
pressure dependence of both simulations. This discrepancy could be due to a different treatment of the density
correction, which at such a high electron energy is very important. Just to play, we have recalculated the deposited
energy with the density correction for 323~hPa in the whole pressure range. Now the resulting energy dependence follows
a similar behavior (see figure~\ref{fig:FLASH_08}).

The results of our detailed simulation for the FLASH experiment are in excellent agreement (within 1\%) with that
obtained from the generic simulation for a sphere of 1.67~cm radius.

The absolute calibration of the optical system of FLASH was carried out by comparison with the measurement of
Rayleigh-scattered light from a narrow beam of a nitrogen laser. They found that efficiency of the optical system for
fluorescence detection is $(3.2\pm 0.25)$\% lower than that for the calibration laser beam, due to the wide-spread
energy deposition from the electron beam. However, our simulation predicts that this effect is negligible in the whole
pressure range for this set-up. Therefore according to our simulation the fluorescence yield reported by FLASH should
be increased by 2\%.

\begin{table}[t!]
\caption{Comparison of absolute values of fluorescence yield from several experiments. Experimental results as given by the
authors are quoted in column~6. Column~9 shows the fluorescence yield in units of photons/MeV resulting from the
normalization to 337~nm, 1013~hPa and 293~K (dry air) using either the assumptions/calculations of the authors (left) or the
results from our simulations (right in bold). See text for details.}

\resizebox{1.0\linewidth}{!}{
\begin{tabular}{*{10}{c}}
\addlinespace\toprule
Experiment                                 &       $\Delta\lambda$      &          $P$          &           $T$        &        $E$       &      Experimental    &        Error         & $I_{337}/I_{\Delta\lambda}$ &       $Y_{337}$      & Correction \\

	                                       &             (nm)           &          (hPa)        &           (K)        &        (MeV)     &        result        &         (\%)         &                             &        (ph/MeV)      &    (\%)    \\

\midrule\midrule
\multirow{5}{*}{Kakimoto~\cite{Kakimoto}}  &             337            &           800         &          288         &        1.4       &       5.7 ph/MeV     &         10           &           1                 & 4.55 / \textbf{4.81} &      6     \\
\cmidrule{2-10}
                                           & \multirow{4}{*}{300 - 400} & \multirow{4}{*}{1013} & \multirow{4}{*}{288} &        1.4       &       3.3 ph/m       & \multirow{4}{*}{10}  &    \multirow{4}{*}{0.279}   & 4.54 / \textbf{4.80} &      6     \\
                                           &                            &                       &                      &      300         &       4.9 ph/m       &                      &                             & 4.44 / \textbf{5.53} &     25     \\
                                           &                            &                       &                      &      650         &       4.4 ph/m       &                      &                             & 3.80 / \textbf{4.85} &     27     \\
                                           &                            &                       &                      &     1000         &       5.0 ph/m       &                      &                             & 4.28 / \textbf{5.51} &     29     \\
 \midrule
 Nagano~\cite{Nagano_04}                   &             337            &          1013         &          293         &        0.85      &       1.021 ph/m     &         13           &           1                 & 5.05 / \textbf{5.35} &      6     \\
 \midrule
 \multirow{2}{*}{Lefeuvre~\cite{Lefeuvre}} & \multirow{2}{*}{300 - 430} & \multirow{2}{*}{1005} & \multirow{2}{*}{296} &        1.1       &       3.95 ph/m      &  \multirow{2}{*}{5}  &    \multirow{2}{*}{0.262}   & 5.15 / \textbf{5.52} &      7     \\
                                           &                            &                       &                      &        1.5       &       4.34 ph/m      &                      &                             & 5.63 / \textbf{6.10} &      8     \\
 \midrule
 \multirow{3}{*}{MACFLY~\cite{MACFLY}}     & \multirow{3}{*}{290 - 440} & \multirow{3}{*}{1013} & \multirow{3}{*}{296} &        1.5       &      17.0 ph/MeV     & \multirow{3}{*}{13}  &    \multirow{3}{*}{0.255}   & 4.32 / \textbf{4.35} &      1     \\
                                           &                            &                       &                      &  $20\cdot 10^3$  &      17.4 ph/MeV     &                      &                             & 4.42 / \textbf{4.34} &     -2     \\
                                           &                            &                       &                      &  $50\cdot 10^3$  &      18.2 ph/MeV     &                      &                             & 4.62 / \textbf{4.53} &     -2     \\
 \midrule
FLASH~\cite{FLASH_08}                      &          300 - 420         &          1013         &          304         & $28.5\cdot 10^3$ &      20.8 ph/MeV     &          7.5         &         0.272               & 5.55 / \textbf{5.65} &      2     \\

 \midrule
 AirLight~\cite{AirLight}                  &             337            &           -           &           -          &      0.2 - 2     & $Y^0=384$ ph/MeV$^a$ &           16         &           1                 & 5.83 / \textbf{5.40} &     -7     \\
 \midrule
 AIRFLY$^b$~\cite{AIRFLY_nagoya}           &             337            &          1013         &          293         &  $120\cdot 10^3$ &       5.6 ph/MeV     &     $\lesssim5\%$    &           1                 &  5.6 / \textbf{ - }  &      -     \\
 \bottomrule\addlinespace%
\multicolumn{10}{l}{\footnotesize $^a$Fluorescence yield at null pressure.}\\
\multicolumn{10}{l}{\footnotesize $^b$Preliminary result using 120~GeV protons.}\\

\end{tabular}
}%
\label{tab:comparison}
\end{table}

The MACFLY experiment~\cite{MACFLY} performed measurements at 20 and 50~GeV using electrons from a linear accelerator
and 1.5~MeV (mean energy) from a $^{90}$Sr source. The energy deposited in the field of view is calculated by the
authors with GEANT4. We have also carried out a detailed simulation with the following features. Primaries colliding
the walls of the chamber or being stopped inside, and so not reaching the trigger system, are rejected in our
simulation. The average energy deposited per electron and unit mass thickness has been obtained from the ratio between
the integrated energy deposition and the total path length of triggering primaries. Statistical uncertainties were
reduced below 0.5\%. The deposited energy obtained from our simulation is very close to that reported by MACFLY.
According to our simulations the fluorescence yield of MACFLY should be increased by 1\% at 1.5~MeV and decreased by
2\% at 20 and 50~GeV. Again the detailed simulation is in good agreement (in this case better than 2\%) with the
generic simulation for $R=10$~cm at energies above 0.5~MeV.

For the experiments of Kakimoto~\etal~\cite{Kakimoto} and Lefeuvre~\etal~\cite{Lefeuvre} we have compared the
predictions of our generic simulation with those assumed by these authors. In the Kakimoto \etal experiment electrons with
a mean energy of 1.4~MeV from a $^{90}$Sr source as well as those from an electron synchrotron with energies of 300,
650 and 1000~MeV were used. Fluorescence light was produced and observed in a 10~cm gap. For the determination of the
fluorescence yield, they assumed full energy deposition in the beam axis. According to our generic simulation, the
deposited energy inside a sphere of 10~cm radius is about 6\%, 25\%, 27\% and 29\% lower than the energy loss for 1.4,
300, 650 and 1000~MeV electrons, respectively. Therefore, the fluorescence yield values at the above energies should be
increased correspondingly by these factors. Results obtained from applying these corrections to measurements of
Kakimoto~\etal are consistent with those obtained in~\cite{Arqueros_Moriond} using an alternative procedure.

In the experiment of Lefeuvre~\etal~\cite{Lefeuvre}, electrons with mean energies of 1.1~MeV (or 1.5~MeV when a higher
energy threshold is applied to the electron detector) from a very intense $^{90}$Sr source (370~MBq) were used.

From our generic simulation, assuming a sphere of 4~cm radius for Lefeuvre \etal experiment, we found that the deposited
energy is 9\% and 10\% lower than the total energy loss at mean electron energies of 1.1 and 1.5~MeV, respectively.
However, electron scattering by the lead shield of this set-up is not included in our simulation, which assumes that
secondaries reaching the walls are absorbed, and thus, it may underestimate the deposited energy. Taking into account
this effect (see~\cite{Rosado_Astropart_Phys} for details), the fluorescence yield of~\cite{Lefeuvre} should be
increased by 7\% (8\%) for 1.1 (1.5 MeV).

\begin{figure}[t]
\centering
\includegraphics[angle=-90,width=0.49\linewidth]{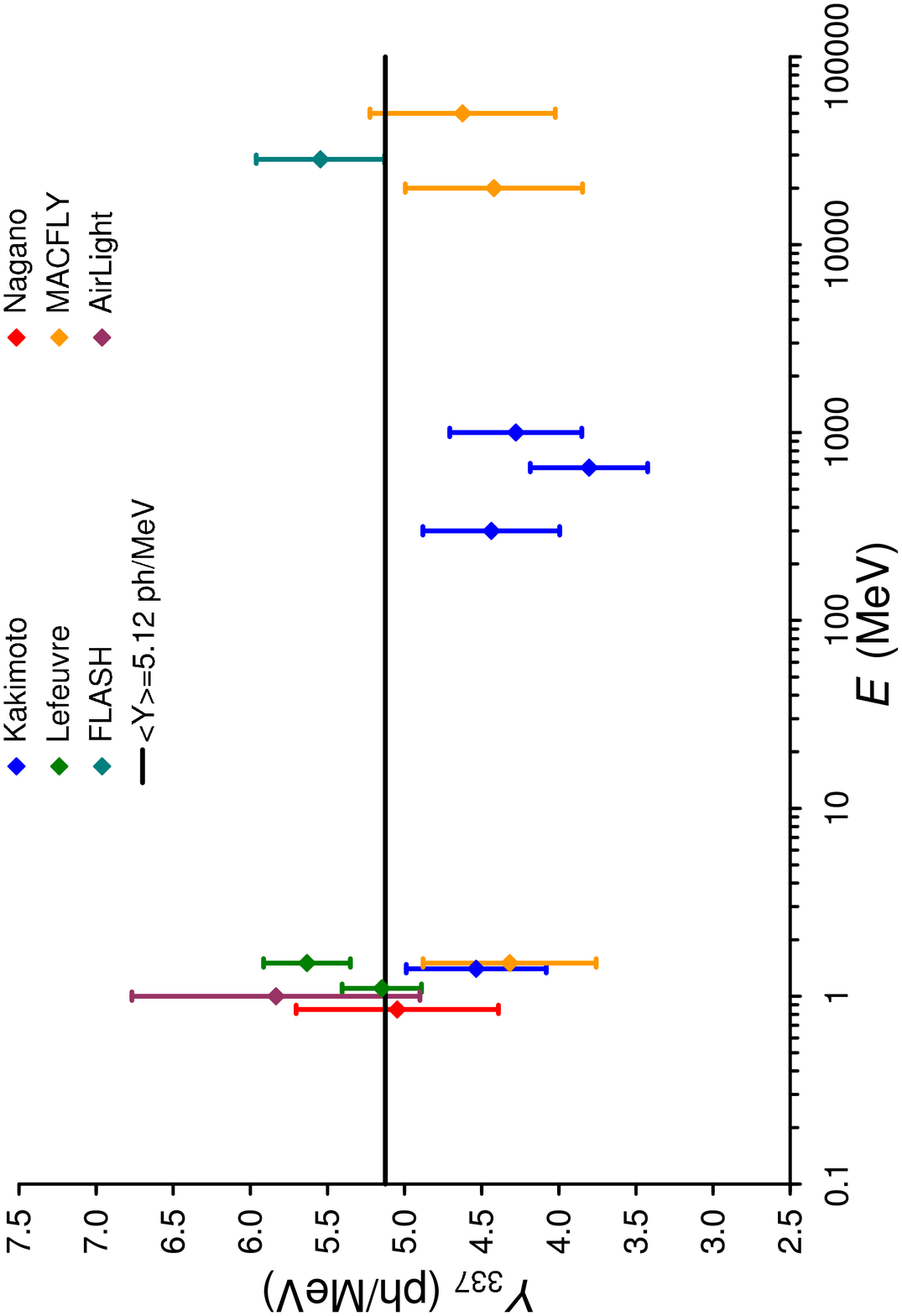}
\includegraphics[angle=-90,width=0.49\linewidth]{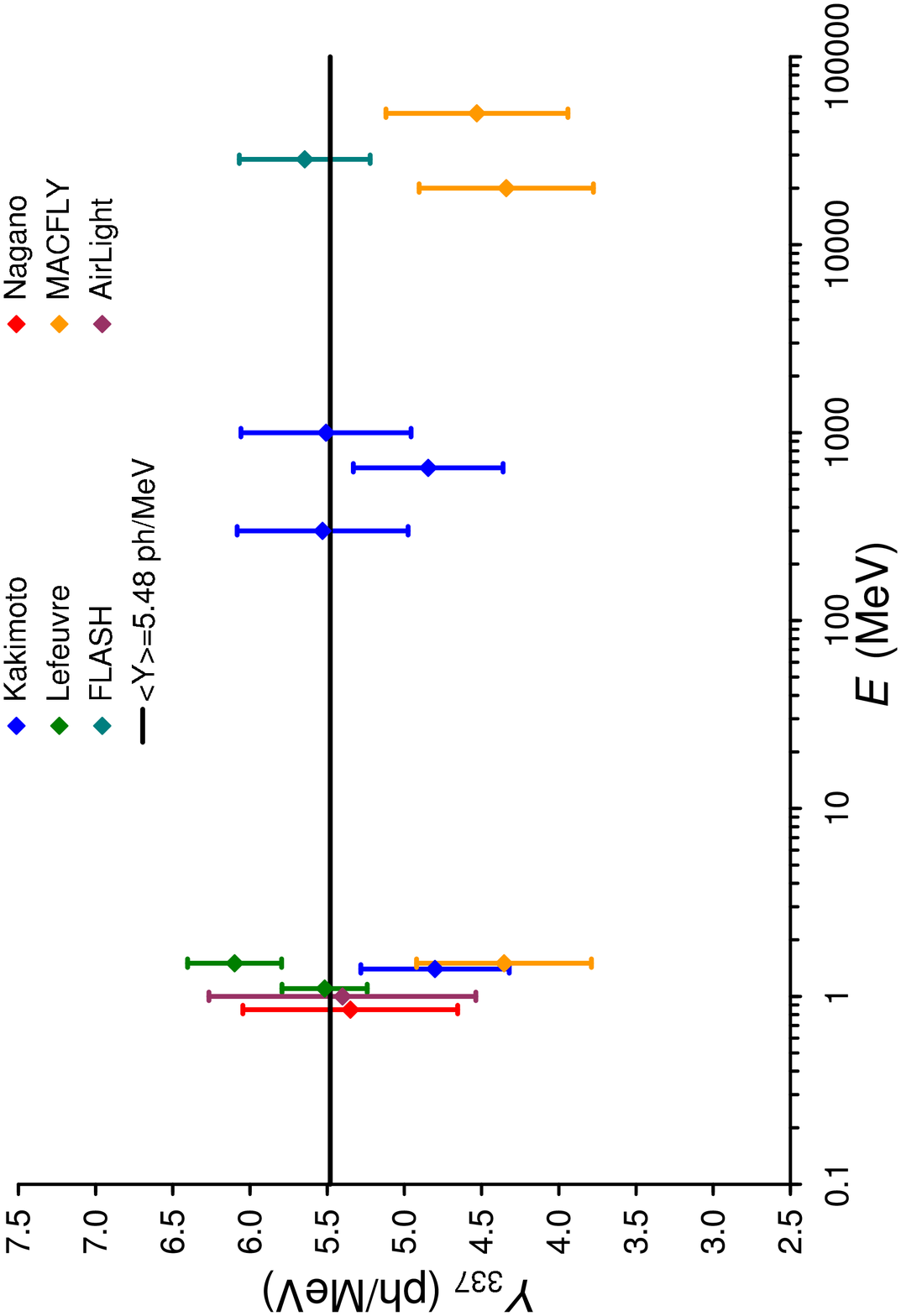}
\caption{Comparison of normalized $Y_{337}$ values as a function of energy for both uncorrected (left) and corrected (right)
values. The horizontal lines represent the corresponding weighted average values.}
\label{FY_E}
\end{figure}

\section{Comparison of fluorescence yield values~\cite{FY_average_value}}

The air-fluorescence yield measurements are listed in table~\ref{tab:comparison} indicating the experimental conditions
(i.e., pressure, temperature, energy, wavelength, units) as well as the uncertainties (column 7). The normalization
factor to get the fluorescence yield for the 337 nm band $I_{337}/I_{\Delta \lambda}$ is shown in column 8. The value
of $Y_{337}$ defined as the fluorescence yield at 1013 hPa and 293 K in ph/MeV units is shown in column 9. Values
inferred from the assumptions (or calculations) on energy deposition and geometrical factors given by the authors are
displayed in left side. In bold the results obtained after applying the corrections proposed in the previous section
(column 10) are shown (column 9 - right).

Several conclusions can be extracted from this table. The uncorrected normalized values are in reasonable agreement
although discrepancies are very often beyond experimental uncertainties. On the other hand, the proposed corrections
are non-negligible (6\% - 29\%) for experiments where the energy deposition is approximated by the electron energy
lost, i.e.,~\cite{Kakimoto,Nagano_04,Lefeuvre}. In general, our simulations are in good agreement with those carried
out by the experiments ($\lesssim $ 2\% for MACFLY and FLASH). In regard to AirLight we have found some discrepancy
(proposed correction of 7\%) but its origin is still under discussion.

Normalized fluorescence yields have been represented as a function of the electron energy in figure~\ref{FY_E}, both
uncorrected (left panel) and corrected (right panel) values. As can be appreciated, measurements are in better
agreement when including our corrections. In addition, the corrected results give more support to the expected energy
independence of the fluorescence yield.

In order to know quantitatively to what extent our corrections favor the agreement between the absolute results
included in this comparison, a statistical analysis has been performed. In the first place, for a given experiment,
results obtained at different energies have been averaged assuming a common systematic error. Then, the average value
of this sample $\langle Y \rangle$ has been calculated weighting the data with the reciprocal of the quoted square
uncertainties shown in column 7 of table~\ref{tab:comparison} (i.e., $w_i=1/\sigma_i^2$). Also the corresponding
variance $\left(\sum_i 1/\sigma_i^2\right)^{-1}$ and the $\chi^2$ statistic divided by the number of degrees of freedom
has been computed. The $\chi^2/{\rm ndf}$ result is larger than 1 for the uncorrected sample, indicating that the
quoted uncertainties are very likely underestimated due to the fact that authors usually do not include any error
contribution from the evaluation of the deposited energy. For the calculation of the uncertainty in the average
fluorescence yield $\sigma_{\langle Y\rangle}$, the variance has been multiplied by $\chi^2/{\rm ndf}$, following the
usual procedure.

\begin{figure}[t]
\centering
\includegraphics[angle=-90,width=0.49\linewidth]{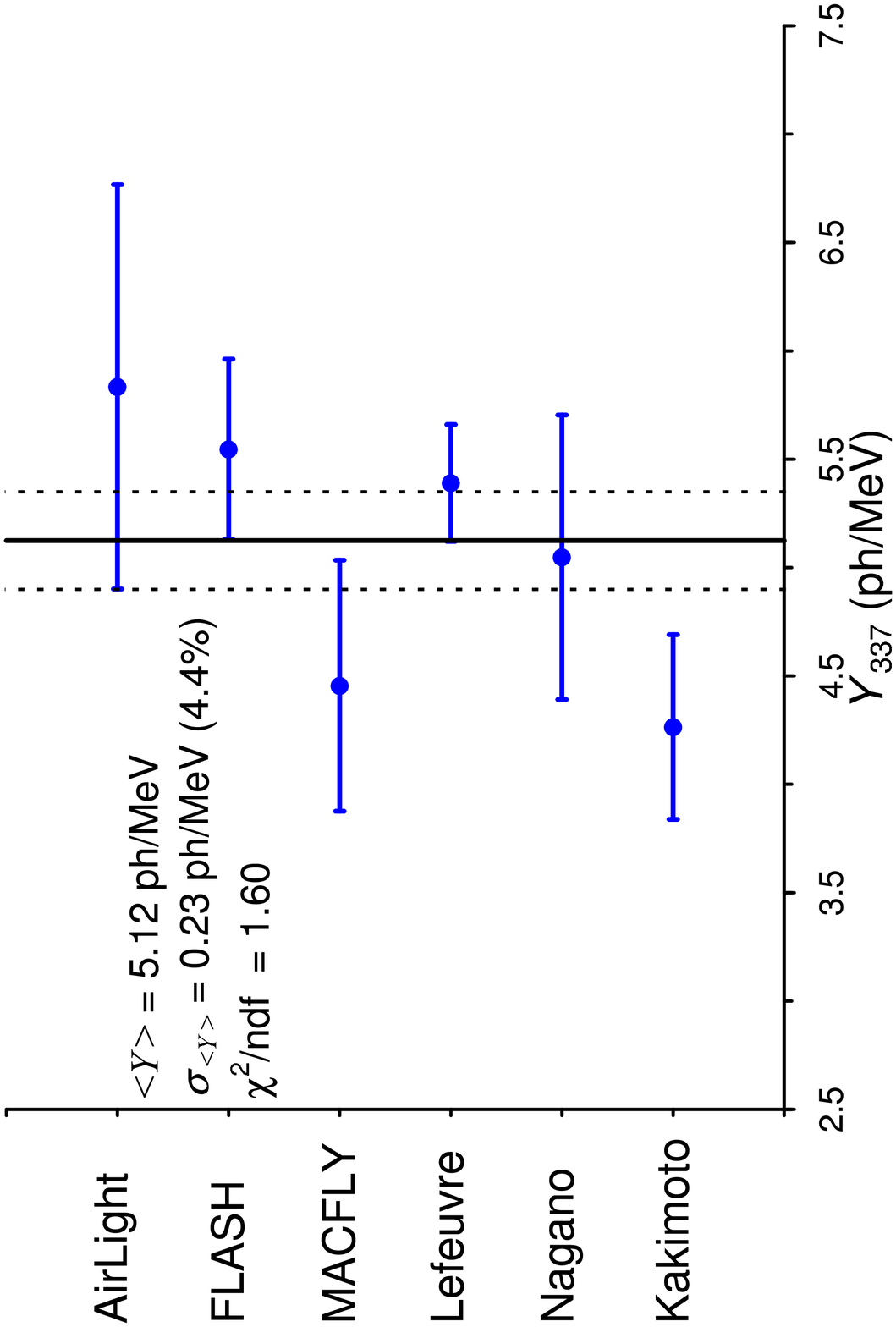}
\includegraphics[angle=-90,width=0.49\linewidth]{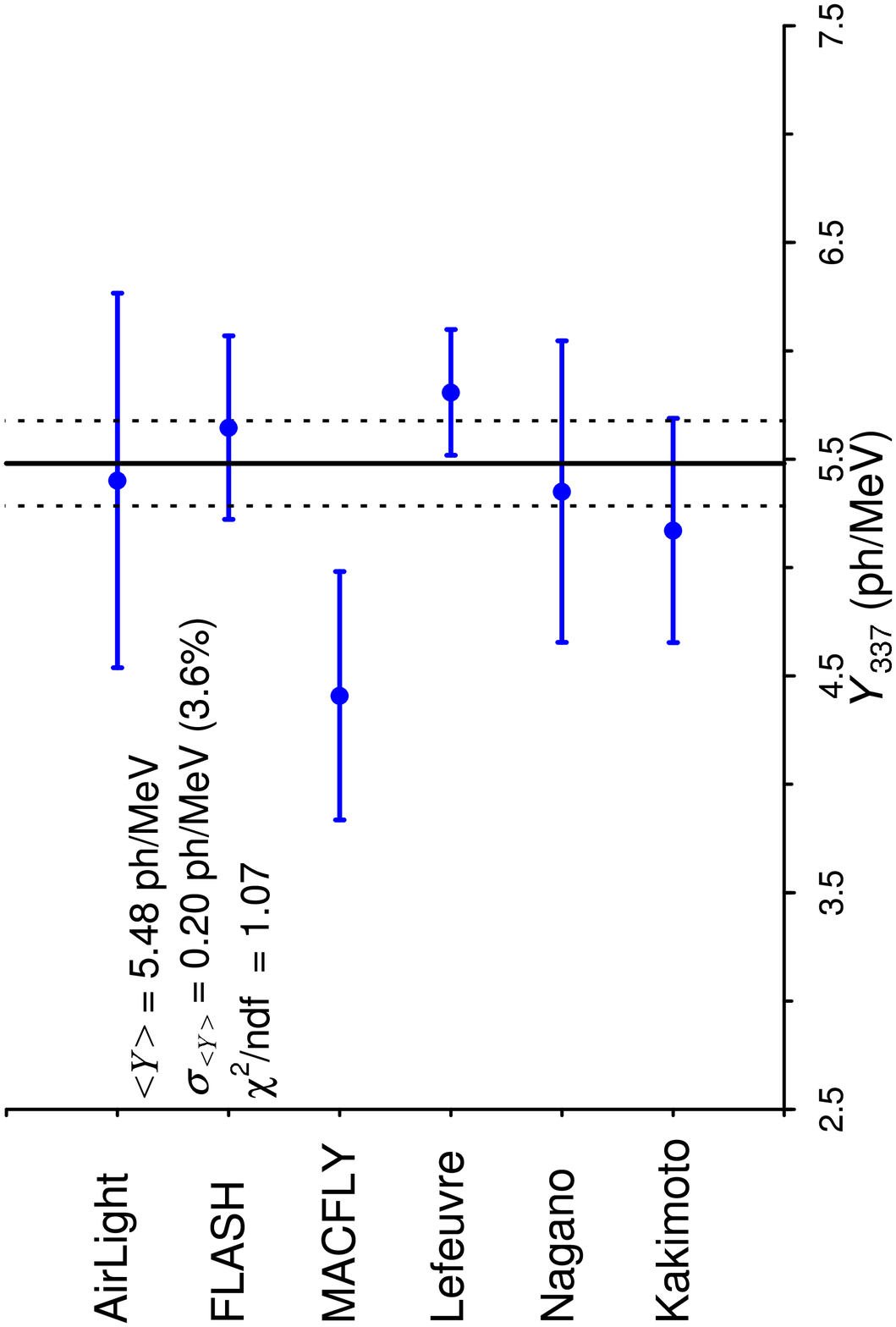}
\caption{Graphical representation of the $Y_{337}$ values at 1013~hPa and 293~K for both uncorrected (left) and corrected
(right) values. The weighted mean $\langle Y\rangle$ (vertical continuous line), the standard deviation of the mean
$\sigma_{\langle Y\rangle}$ (half of interval between dashed lines) and the $\chi^2$ statistic normalized by the number of
degrees of freedom are shown in the legends. Better agreement and consistency of the data sample are found
in the right plot.}
\label{FY_av}
\end{figure}

The results shown in figure~\ref{FY_av} for both uncorrected (left panel) and corrected (right panel) values indicate
that our corrections lead to a more consistent data sample suggesting that they do improve the determination of the
deposited energy for the different experiments.

We have checked~\cite{FY_average_value} that the weighting procedure has no significant effect on the final result. In
addition, removing each value while keeping the remaining ones in the corrected sample does not significantly change
the weighted average. The corresponding results range from 5.61~ph/MeV (removing MACFLY) to 5.23~ph/MeV (removing
Lefeuvre~\etal). A detailed analysis of these features~\cite{FY_average_value} including the effect of a weak energy
dependence of the fluorescence yield in the $0.1-10$~MeV range, predicted by our simulations, led us to an average
value of $Y_{337}=5.45$~ph/MeV with a conservative estimated error of $5\%$. According to the comparison of our
simulation result on energy deposition with GEANT4, a small systematic uncertainty of below 2\% should be added,
although it does not affect the $\chi^2$ value of the corrected sample.

The recent absolute measurement of the AIRFLY collaboration~\cite{AIRFLY_nagoya} yields $Y_{337}=5.6$~ph/MeV with an
uncertainty of $\lesssim5\%$ (still preliminary), which is fully compatible with the above value. If this new result is
included in the average, then a weighted mean value of 5.52~ph/MeV is obtained with an uncertainty of $\lesssim5\%$.

As already mentioned, for the comparison presented here we have normalized the air-fluorescence yield measurements to
its value for the 337~nm band at 1013~hPa and 293~K. However, in some occasions it might be more convenient to use the
integral of the fluorescence yield in a wider spectral range and/or other pressure and temperature conditions. The
conversion can be easily done following the procedure described in detail in~\cite{Arqueros_NJP}. For instance, the
above average value would be of 20.1~ph/MeV ($\pm 5\%$) for the $300-420$~nm spectral range at the same reference
pressure and temperature, which would become 20.3~ph/MeV if the measurement of AIRFLY is included.

Our simulation can also provide a theoretical value of the air-fluorescence yield. Unfortunately, the evaluation of the
fluorescence emission cannot be very precise due to the large uncertainties in the relevant molecular parameters.
Therefore, we expect a large uncertainty in such a calculation of the fluorescence yield, which we estimated to be
about 25\%~\cite{Arqueros_NJP}. Nevertheless, a result of $Y_{337}$ = 6.3~ph/MeV (using the quenching parameter
of~\cite{AIRFLY_P}) has been found, which is consistent with the experimental ones, providing a valuable theoretical
support to these measurements.

\section{Conclusions}

Available measurements of the absolute air-fluorescence yield have been normalized to common conditions (1013~hPa,
293~K, 337~nm) and units (ph/MeV). According to the simulations presented in this work, experimental results obtained
neglecting the energy deposited by secondary electrons outside the field of view of the optical system have to be
corrected by non-negligible factors ranging from 6 to 29\%. Our evaluation of energy deposition is, in general, in
agreement with that reported by those experiments which carried out a detailed simulation. The corrections to the
measurements of the absolute air-fluorescence yield proposed here increase significantly the compatibility of
experimental results. An average value of $Y_{337}=5.45$~ph/MeV with a 5\% uncertainty has been obtained. If the
absolute fluorescence yield and error of AIRFLY are confirmed, a consensus on this important parameter with an
uncertainty below the 5\% level could be reached with high reliability.

\section*{Acknowledgements}
This work has been supported by the Spanish Ministerio de Ciencia e Innovacion (FPA2009-07772 and CONSOLIDER CPAN
CSD2007-42) and ``Comunidad de Madrid" (Ref.: 910600). J.~Rosado acknowledges a PhD grant from ``Universidad
Complutense de Madrid". The authors thank our colleagues of the Auger collaboration for fruitful discussions and
comments on this work.

\end{document}